\documentclass{lmcs} 


\usepackage{tabularx}
\usepackage{makecell}
\usepackage{array}
\usepackage{fancyvrb}
\usepackage{textgreek}
\usepackage{orcidlink}
\usepackage{booktabs}
\usepackage{listings}
\usepackage[utf8]{inputenc}
\usepackage[T1]{fontenc}
\usepackage{pifont}
\usepackage{hyperref}
\usepackage{alltt}
\usepackage{listings}
\theoremstyle{plain} 

\begin{document}

\title[Instructions]{Formal Modeling and Verification of Grover's Algorithm}
\thanks{SUN and SHI are co-first authors with equal contributions.}
\thanks{WANG is the corresponding author.}

\author[H.~Sun]{Hongxia Sun\lmcsorcid{0009-0002-1513-1823}}[a]
\author[Z.~Shi]{Zhiping Shi\lmcsorcid{0000-0002-3562-8602}}[b]
\author[S.~Chen]{Shanyan Chen\lmcsorcid{0009-0005-0837-5215}}[c]
\author[G.~Wang]{Guohui Wang\lmcsorcid{0000-0002-3176-317X}}[a]
\author[X.~Li]{Ximeng Li\lmcsorcid{0009-0001-8991-2756}}[b]
\author[Y.~Guan]{Yong Guan\lmcsorcid{0000-0002-2373-2779}}[d]
\author[Q.~Zhang]{Qianying~Zhang\lmcsorcid{0000-0002-3246-9474}}[a]
\author[Z.~Shao]{Zhenzhou~Shao\lmcsorcid{0000-0002-9166-9468}}[a]
\address{College of Information Engineering, Capital Normal University, Beijing}	
\email{hxsun@cnu.edu.cn; ghwang@cnu.edu.cn; qyzhang@cnu.edu.cn; zshao@cnu.edu.cn}  

\address{Beijing Key Laboratory of Electronic System Reliability Technology, Capital Normal University, Beijing}	
\email{shizp@cnu.edu.cn; lixm@cnu.edu.cn}  

\address{School of Mathematical Sciences, Capital Normal University, Beijing}	
\email{sychen@cnu.edu.cn} 

\address{International Science and Technology Cooperation Base of Electronic System Reliability and Mathematical Interdisciplinary, Capital Normal University, Beijing}	
\email{guanyong@cnu.edu.cn} 




\begin{abstract}
  \noindent Grover's algorithm relies on the superposition and interference of quantum
mechanics, which is more efficient than classical computing in specific tasks such
as searching an unsorted database. Due to the high complexity of quantum mechanics, the correctness of quantum algorithms is difficult to guarantee through
traditional simulation methods. By contrast, the fundamental concepts and mathematical structure of Grover's algorithm can be formalized into logical expressions
and verified by higher-order logical reasoning. In this paper, we formally model
and verify Grover's algorithm in the HOL Light theorem prover. We focus on
proving key properties such as the unitarity of its oracle and diffusion operators, the monotonicity of the success probability with respect to the number of
iterations, and an exact expression for the optimal iteration count. By analyzing
a concrete application to integer factorization, we demonstrate the practicality
and prospects of our work.
\end{abstract}

\maketitle

\section*{Introduction}\label{S:one}

  Quantum computing, as a transformative paradigm, demonstrates processing capabilities far beyond those of classical computing for specific problems~\cite{17MitraQuantumcryptography, zhou2020quantumapproximate, pyrkov2023quantum}. Represented by Grover's algorithm, quantum search technology stands as a milestone in quantum computing~\cite{yao2022feedback, chang2020Thesecond}, significantly reducing the time complexity of unsorted database search from the classical $O(N)$ to $O(\sqrt N)$~\cite{Grover}. Owing to this superior complexity,
Grover’s algorithm has great theoretical importance. Furthermore, it demonstrates broad application prospects in various practical domains such as combinatorial optimization~\cite{zhang2024quantum, liu2025solving}, cryptanalysis~\cite{song2022speedy,yoon2023quantum}, and database retrieval~\cite{sun2024quantum, armenakas2020application}.

However, as quantum algorithms progressively move toward engineering implementation, ensuring their correctness has become a major bottleneck restricting the reliability and development of quantum computing~\cite{lewis2023formal}. Inherent quantum properties such as superposition, measurement collapse, and probabilistic output make it difficult for traditional testing methods to cover all possible execution paths~\cite{chareton2023formal}. Meanwhile, manual mathematical derivation and verification methods lack scalability and are highly error-prone~\cite{SSJDE271DC03D10A806731F63320BFF6F1D4}.

Against this backdrop, formal verification methods based on higher-order logic theorem provers demonstrate unique value. This approach enables machine-assisted, fully reliable verification of algorithmic properties and functions within a strict mathematical axiomatic framework. Through rigorous mathematical reasoning, formal verification can fundamentally identify and eliminate system vulnerabilities and security risks at the algorithm design level~\cite{hasan2015formalverificationmethods}, thereby providing dependable assurance for the correctness of quantum algorithms. In recent years, numerous researchers have achieved significant progress in quantum computing by utilizing theorem proving tools.
Liu et al.~\cite{liu2019formal} formalized the syntax and semantics of quantum programs in Isabelle/HOL, wrote down the rules of quantum Hoare logic, and verified the soundness and completeness of the deduction system for partial correctness of quantum programs.
Lin et al.~\cite{lin2024parallel} proposed a parallel and distributed quantum SAT solver that exploits entanglement and teleportation. By introducing auxiliary entangled qubits, they reduced the time complexity of each Grover iteration from $O(N)$ to $O(1)$.
Bordg et al.~\cite{bordg2021certified} formally verified several quantum algorithms, including quantum teleportation and Deutsch's algorithm, in Isabelle/HOL. Hietala et al.~\cite{hietala2021proving} independently verified quantum phase estimation using the SQIR languages. Shi et al.~\cite{shi2021symbolic} conducted symbolic verification of both the Deutsch-Jozsa algorithm and quantum teleportation using Coq. These results demonstrate the applicability of formal verification of quantum algorithms, providing an important reference and practical foundation for the formal verification of Grover's algorithm in our paper.

In addition, Chareton et al.~\cite{chareton2021automated} proposed the QBRICKS verification framework, which adopts parameterized paths and circuit descriptions to represent quantum circuits in a functional programming style effectively. However, this representation is measurement-free. Separately, Shi et al.~\cite{shi2024formalizing} formalized a classical-quantum hybrid language in Coq and verified the final state correctness of Grover's algorithm. Still, their work is confined to the two-qubit case and does not cover aspects such as the derivation of the optimal iteration number or the analysis of the success probability (the probability of finding correct solutions) distribution with respect to the number of iterations.

Overcoming the aforementioned limitations in measurement semantics and verification scale, our work enables a deeper and more comprehensive formal verification of Grover's algorithm. This paper formally verifies each step of Grover's algorithm in HOL Light\footnote{The code is available at \url{ https://github.com/shx-02/Grover-algorithm}}, starting from the algebraic definitions of quantum states and operators. We provide an inductive proof of its correctness for an arbitrary number of qubits. Additionally, this paper provides a case study on integer factorization to demonstrate the practical application value of Grover's algorithm. Our main contributions include
\begin{itemize}
    \item Development of a complete formal model for Grover's algorithm, incorporating formal definitions of its quantum states, oracle, and diffusion operator.
\end{itemize}
\begin{itemize}
    \item Verification of key algorithmic properties, including the monotonicity of success probability and the formula for the optimal iteration count.
\end{itemize}
\begin{itemize}
    \item Demonstration of the framework's practicality through an application to integer factorization.
\end{itemize}

\subsection*{Paper Structure} Section \ref{section 2} provides the formalization of the basic theory of Grover's algorithm. Section \ref{section 3} constructs a formal model of the quantum evolution from the initial quantum state to the final measurement. Section \ref{section 4} presents the verification of core properties, such as the optimal number of iterations. Section \ref{section 5} introduces a brief case study on integer factorization for illustration. Finally, we summarize the paper in Section \ref{section 6}.

\section{Formalization of Basic Theory}
\label{section 2}
  The mathematical description of Grover's algorithm is based on linear algebra and quantum mechanical principles. This section systematically reviews the requisite theory, laying a solid groundwork for the subsequent formal verification and in-depth applications.

\subsection{Linear Algebra}
\indent The mathematical framework of quantum mechanics is formulated on complex vector spaces. Vectors (quantum states) in this space are denoted using the Dirac notation: a ket $|\varphi\rangle$ represents a column vector, while a bra $\left\langle\varphi\right|$ represents its dual vector (conjugate transpose). The inner product on this complex vector space $V$ is a mapping
\begin{equation}
    \label{eqinner}
    \langle \cdot | \cdot \rangle : V \times V \to \mathbb{C}
\end{equation}
which must satisfy the properties of conjugate symmetry, linearity, and positive definiteness.

\indent A complex matrix is a linear transformation on a complex vector space. Among them, matrices with specific structures and properties, such as the identity matrix and diagonal matrices, can significantly simplify the verification of quantum gate unitarity. Table \ref{tab1properties} shows the key properties of these two types of matrices.
\begin{table}[!ht]
\caption{Key properties of matrices}
  \label{tab1properties}
\begin{tabularx}{\textwidth}{>{\centering\arraybackslash\hsize=0.7\hsize}X>{\centering\arraybackslash\hsize=1.1\hsize}X>{\centering\arraybackslash\hsize=1.2\hsize}X}
\hline
\toprule
    property  & formula  & pseudo-code\\
    \midrule
    self-adjointness&$I=I^\dagger$ &\texttt{id\_cmatrix = hermitian\_matrix id\_cmatrix} \\
    diagonality&$I=diag(1,\cdots,1)$	&\texttt{diagonal\_cmatrix id\_cmatrix}\\
left-multiplicative identity&$IA=A,Iq=q$&\texttt{id\_cmatrix ** A = A,
id\_cmatrix ** q = q}\\
right-multiplicative identity&$AI=A$&\texttt{A ** id\_cmatrix = A}\\
symmetry&$D=D^T$&\texttt{diagonal\_cmatrix A $\implies$A = ctransp A}\\
diagonal-matrix multiplication&	
  $A=\mathrm{diag}(a_{1},\dots,a_{n}) ~\land
  B=\mathrm{diag}(b_{1},\dots,b_{n})\Longrightarrow
  AB=\mathrm{diag}(a_{1}b_{1},\dots,a_{n}b_{n})$
 &\texttt{diagonal\_cmatrix A $\land$ diagonal\_cmatrix B $\implies$A ** B = $\lambda$ij.~A$_{ij}$B$_{ij}$}\\
    \bottomrule
\hline
\end{tabularx}
\end{table}
\begin{defi}\label{de1.1}
  Hermitian Conjugate
\end{defi}
The conjugate transpose operation of a complex matrix $A$ is called the Hermitian conjugate, denoted as $A^\dagger$, i.e., $A^\dagger = (A^*)^T$.
\begin{lstlisting}[language=C,mathescape] 
let hermitian_matrix = new_definition
`(hermitian_matrix:$\mathbb{C}^{2^N\times 2^M}\rightarrow \mathbb{C}^{2^M\times 2^N})$ A=$\lambda$ij.cnj(A$_{ji}$)`
\end{lstlisting}
Here, $A^*$ denotes the complex conjugate of the elements of $A$.

This operation is fundamental to defining unitary matrices and is a key to verifying the legality of quantum state evolution. Next, we use the Hermitian conjugate to define a unitary matrix.
\begin{defi}\label{de1.2}
 Unitary Matrix
\end{defi}
A unitary matrix is a complex square matrix whose conjugate transpose is precisely its inverse, as expressed mathematically in Equation \ref{unitary}.
\begin{equation}
    \label{unitary}
    U^\dagger U = UU^\dagger = I
\end{equation}
\begin{lstlisting}[language=C,mathescape] 
let unitary_matrix = new_definition
`unitary_matrix(U:$\mathbb{C}^{2^N\times 2^N}$) $\Longleftrightarrow$ 
  hermitian_matrix U ** U = id_cmatrix $\land$ U ** hermitian_matrix U = id_cmatrix`
\end{lstlisting}

To facilitate the verification of probability conservation during state evolution, we further derive that the column (or row) vectors of a unitary matrix are mutually orthogonal and normalized.
\begin{thm}\label{de1.3}
 Orthogonality and Normalization of Unitary Matrix
\end{thm}
If $U$ is a unitary matrix,  then its column vectors $u_1,u_2,...,u_n$, satisfy the following properties: (i) 
normalization, if $i = j$, then $u_i^\dagger u_j = 1$, and (ii) orthogonality, if $i\ne j$, then $u_i^\dagger u_j = 0$.
Their corresponding expression using the Kronecker delta function $\delta$ is given in Equation \ref{eq5}.
\begin{equation}
    \label{eq5}
    \sum_{i=1}^{n} U_{ix} \overline{U_{iy}} = \delta_{xy} =
    \begin{cases} 
1,&\text{if } x = y, \\
0,&\text{otherwise.}
\end{cases}
\end{equation}
\begin{lstlisting}[language=C,mathescape] 
$\forall$A:$\mathbb{C}^{2^N\times 2^N}$. unitary_matrix A $\implies$ ($\forall$x y. 1 $\leq$ x $\land$ x $\leq 2^N \land$ 1 $\leq$ y $\land$ y $\leq 2^N$) $\implies$ 
 vsum(1..$2^N$)($\lambda$i. A$_{ix}$ * cnj(A$_{iy}$)) = if x = y then Cx(&1) else Cx(&0))
\end{lstlisting}
Here, \texttt{vsum} is for summing the elements of a vector, corresponding to the \(\Sigma\) in Equation \ref{eq5}. \texttt{Cx(\&1)} represents converting the real 1 into a complex number.
\begin{thm} 
\label{de1.4}
Closure of Unitary Matrices Under Multiplication
\end{thm}

The product of two unitary matrices is also a unitary matrix.
\begin{lstlisting}[language=C,mathescape]
 $\forall$A B:$\mathbb{C}^{2^N\times 2^N}$. unitary_matrix A $\land$ unitary_matrix B $\implies$ unitary_matrix (A ** B)
\end{lstlisting}

The tensor product is a method of combining vector spaces together to form a larger space. In quantum computing, we frequently need to compute the tensor product of multiple single-qubit gates (i.e., 2-dimensional unitary matrices) to construct multi-qubit gates. Therefore, this paper formally defines a function that supports computing the tensor product of any number of 2-dimensional matrices.
\begin{defi}
    \label{de1.5}
    Matrix Tensor Product
\end{defi}

Let $A_1, A_2, \dots, A_k$ be $2 \times 2$ complex matrices. Their tensor product $A_1 \otimes A_2 \otimes \dots \otimes A_k$ is the $2^k \times 2^k$ matrix defined by:

\[
(A_1 \otimes \cdots \otimes A_k)_{ij} = \prod_{l=0}^{k-1} \left( A_{k-l} \right)_{p_l, q_l}
\]
where for $l = 0, \dots, k-1$:
\[
p_l = \left\lfloor \frac{i-1}{2^l} \right\rfloor \bmod 2 + 1, \quad 
q_l = \left\lfloor \frac{j-1}{2^l} \right\rfloor \bmod 2 + 1
\]
with $\lfloor x \rfloor$ denoting the floor function, and $(A_{k-l})_{p_l q_l}$ denotes the element at row $p_l$, column $q_l$ of matrix $A_{k-l}$.
\begin{lstlisting}[language=C,mathescape]
let tensor_n2 = new_definition
`$\forall$m:($\mathbb{C}^{2\times 2}$) list. tensor_n2 m = $\lambda$i j. cproduct (0.. LENGTH m - 1) ($\lambda$k. EL k 
 (REVERSE m)$_{(\lfloor (i-1)/2^k \rfloor \bmod 2+1) (\lfloor (j-1)/2^k \rfloor \bmod 2+1)}$)`
\end{lstlisting}
\begin{defi}
\label{de1.6}
Matrix List Generator
\end{defi}

An $n$-length matrix list is a finite sequence $(A_0, A_1, \dots, A_{n-1})$, where each $A_i$ is a $2\times 2$ complex matrix.
\begin{lstlisting}[language=C,mathescape]
let cmatrix_list = new_definition
`$\forall$n:$\mathbb{N}$ f:$\mathbb{N}\rightarrow \mathbb{C}^{2\times 2}$. (cmatrix_list f n):($\mathbb{C}^{2\times 2}$) list = MAP f (list_of_seq ($\lambda$k. k) n)`
\end{lstlisting}
This supports the generation of a list of any combination of $2 \times 2$ matrices, where \texttt{f} is the matrix generator function and \texttt{n} is the number of matrices.
\subsection{Fundamentals of Quantum Mechanics}
The four cornerstones of quantum mechanics-states, gates, evolution, and measurement-collectively form the formal language of quantum computing. Quantum states encode information, quantum gates implement unitary transformations of information, unitary evolution ensures the conservation of probability, and measurement accomplishes the transition from quantum to classical information. Each component is examined in turn below to build a minimal, runnable set of mechanical rules.

In quantum mechanics, a pure state must satisfy the normalization condition. Therefore, we introduce the general predicate \texttt{is\_qstate} to assert that a complex vector has unit squared norm, and then use this predicate to define a new quantum-state type.
\begin{defi}
    \label{de1.7}
    Quantum State
\end{defi}

An $n$-qubit pure state is a \( 2^n \)-dimensional complex vector $|\varphi\rangle$ satisfying \(\langle\varphi|\varphi\rangle=1\). 
\begin{lstlisting}[language=C,mathescape]
 let is_qstate = new_definition
 `is_qstate (q:$\mathbb{C}^M$) $\Leftrightarrow$ cnorm2 q = &1`
 let qstate_tybij = prove
     let th = prove
   (`$\exists$q:$\mathbb{C}^{2^N}$. is_qstate q`) in new_type_definition 
 "qstate" ("mk_qstate","dest_qstate") th
\end{lstlisting}
Here, \texttt{cnorm2} represents the squared norm of a complex vector $\|\cdot\|^2$. The function \texttt{mk\_qstate} maps a complex vector to a quantum state, while \texttt{dest\_qstate} extracts the original complex vector from a quantum state.

The evolution of a quantum system is implemented through linear transformations described by quantum gates. Based on the number of qubits they act upon, quantum gates are categorized into single-qubit gates and multi-qubit gates. Among various single-qubit gates, the Hadamard gate (H-gate) holds a fundamental role, serving as a key component for enabling quantum parallelism and quantum interference phenomena. Its definition and unitarity verification are provided below.
\begin{defi}
    \label{de1.8}
    Hadamard Gate
\end{defi}

The single-qubit Hadamard gate is a $2\times2$ unitary matrix represented as:
   \[ H=\frac{1}{\sqrt{2}}\begin{bmatrix}1&1
 \\1&-1\end{bmatrix}\]
\begin{lstlisting}[language=C,mathescape] 
let hadamard = new_definition
  `hadamard: $\mathbb{C}^{2 \times 2}$ = $\lambda$i j.
     if i = 1 $\land$ j = 1 then Cx(&1 / sqrt(&2)) else
     if i = 1 $\land$ j = 2 then Cx(&1 / sqrt(&2)) else
     if i = 2 $\land$ j = 1 then Cx(&1 / sqrt(&2)) else
     if i = 2 $\land$ j = 2 then $-$Cx(&1 / sqrt(&2)) else Cx(&0)`
\end{lstlisting}
\begin{thm}
    \label{de1.9}
    Unitarity of the Hadamard Gate
\end{thm}

The Hadamard gate is unitary. 
\begin{lstlisting}[language=C,mathescape] 
 unitary_matrix hadamard
\end{lstlisting}
\begin{defi}
    \label{de1.10}
    State Evolution
\end{defi}

The evolution of a closed quantum system is described by a unitary matrix $U$. The state of the system at the initial time, denoted as \( |\varphi\rangle \), and the state after evolution, denoted as \( |\varphi'\rangle \), are related by \(|\varphi'\rangle=U|\varphi\rangle \). 
\begin{lstlisting}[language=C,mathescape] 
let cmatrix_qstate_mul = new_definition
 `$\forall$q:(N)state A:$\mathbb{C}^{2^N \times 2^M}$. A ** q  = mk_qstate ($\lambda$i. vsum (1..$2^N$) ($\lambda$j. A$_{ij}$ * 
 (dest_qstate q)$_j$))`
\end{lstlisting}
\begin{thm}
    \label{de1.11}
    Conservation of Probability
\end{thm}

Let \( |\psi\rangle \) be a normalized quantum state satisfying \( \langle\psi|\psi\rangle = 1 \), 
and let \( U \) be a unitary matrix. Then the evolved state \( |\psi'\rangle = U |\psi\rangle \) satisfies
$\langle\psi'|\psi'\rangle = 1$.
\begin{lstlisting}[language=C,mathescape] 
 $\forall$A:$\mathbb{C}^{2^N \times 2^N}$ q:(N)state. unitary_matrix A $\implies$
  cnorm2(($\lambda$i. vsum (1..$2^N$)($\lambda$j. A$_{ij}$ * (dest_qstate q)$_j$)):$\mathbb{C}^{2^N}$) = &1
\end{lstlisting}

Unitary evolution completely describes the deterministic dynamics of a quantum system. However, it cannot directly account for the probabilistic events inherent in experimental observations. To formally describe such events and their probabilities, we introduce projection operators as the mathematical embodiment of the measurement postulate.
\begin{defi}
    \label{de1.12}
    Projection Operator
\end{defi}

A projective measurement is characterized by a set of projection operators $\{P_m\}$, satisfying $P_m = |m\rangle\langle m|$, where each basis vector $|m\rangle$ corresponds to a measurement outcome $m$.
\begin{lstlisting}[language=C,mathescape] 
let project = new_definition
  `(project:(N)qstate $\rightarrow \mathbb{C}^{2^N \times 2^N}$) v = qouter v v`
\end{lstlisting}
Here, \texttt{qouter} generates the outer product of $|v\rangle$. 

These projection operators satisfy self-adjointness, idempotence, and completeness:
\begin{thm}
    \label{de1.13}
    Self-Adjointness
\end{thm}

A projection operator $P$ is self-adjoint: $P=P
^\dagger$.
\begin{lstlisting}[language=C,mathescape] 
project (v:(N)qstate) = hermitian_matrix (project v)
\end{lstlisting}
\begin{thm}
    \label{de1.14}
    Idempotence
\end{thm}

A projection operator $P$ is idempotent: $P=P^2$
\begin{lstlisting}[language=C,mathescape] 
(project (v:(N)qstate)) ** (project v) = project v
\end{lstlisting}
\begin{thm}
    \label{de1.15}
    Completeness
\end{thm}

For the set of projection operators $\{P_m\}$, the completeness relation is always satisfied $\sum_m P_m = I$.
\begin{lstlisting}[language=C,mathescape] 
 cmat_sum (1..$2^N$) ($\lambda$i. project (qbasis i:(N)qstate)) = id_cmatrix:$\mathbb{C}^{2^N \times 2^N}$
\end{lstlisting}
Here, \texttt{cmat\_sum} adds up all projection operators \texttt{project (qbasis i)} for $i$ in the range $1\cdots2^N$, yielding the identity matrix.
\begin{defi}
    \label{de1.16}
    Measurement
\end{defi}

When the state to be measured is $|\varphi\rangle$, the probability of obtaining outcome $m$ is $
p(m) = \langle \varphi|P_m|\varphi\rangle$.
\begin{lstlisting}[language=C,mathescape] 
 let measurement = new_definition
 `$\forall$x y:(N)qstate. measurement x y = cnorm2(apply_cmatrix (project x) y)`
\end{lstlisting}
Here, \texttt{apply\_cmatrix} applies the projection operator $|x\rangle\langle x|$ to the quantum state $|y\rangle$, yielding a new complex vector, 
whose squared norm gives the probability of obtaining outcome $x$.
\section{Formal Modeling of Grover's Algorithm}
\label{section 3}
The algorithm is described as follows. In an unsorted database of size $N = 2^n$, only one target item $x'$ satisfies $f(x') = 1$, while all other items give $f(x) = 0$. The task is to identify $x'$ with as few quantum queries as possible.

The algorithm begins by applying the Hadamard (H) gate to the zero state, creating a uniform superposition. This uniform superposition is then fed into the process $G$, which performs $k = O(\sqrt N)$ successive iterations, gradually amplifying the amplitude of the target state. Finally, the target state is measured to obtain the search result. Here, $N = 2^n$, where $n$ represents the number of qubits.

\begin{figure}[h]
  \centering
  \includegraphics[width=0.8\textwidth]{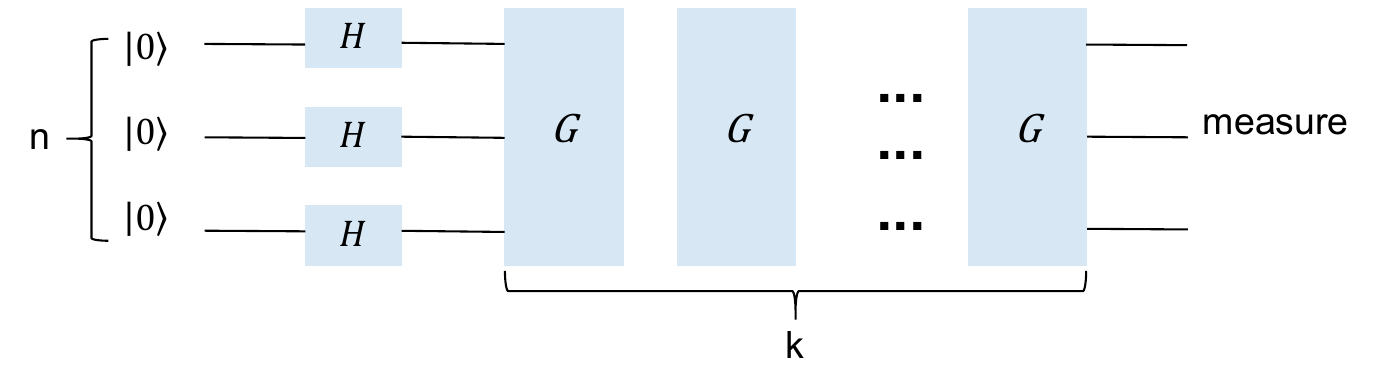}
  \caption{Quantum circuit diagram of Grover's algorithm}
  \label{figure1}
\end{figure}

According to the principles of Grover's algorithm, the solution process can be divided into three modules: the Initialization Module, the Grover Iteration Module, and the Measurement Module. The Grover Iteration Module itself consists of an Oracle Module ($U_f$) and a Diffusion Operator Module (D). Figure \ref{figure1} shows the quantum circuit diagram of Grover's algorithm.

\noindent\textbf{(1) Quantum State Initialization}

Initialize $n$ single qubits to \( |0\rangle^{\otimes n} \), as shown in Equation \ref{eq7}.
 \begin{equation}
     \label{eq7}
|0\rangle^{\otimes n} = \underbrace{|0\rangle \otimes |0\rangle \otimes \cdots \otimes |0\rangle}_{n \text{ times}} = \begin{pmatrix} 1 \\ 0 \\ \vdots \\ 0 \end{pmatrix}\in \mathbb{C}^{2^{n}}
 \end{equation} 
Equation \ref{eq7} can be formally expressed as Definition~\ref{de2.1}.
\begin{defi}
    \label{de2.1}
    Initial State
\end{defi}

\begin{lstlisting}[language=C,mathescape] 
 let zero_qstate = new_definition
 `zero_qstate:(N)qstate = mk_qstate ($\lambda$i.if i = 1 then Cx(&1) else Cx(&0))`
\end{lstlisting}
\textbf{(2) Application of $H^{\otimes n}$}

Apply the H gate individually to each of the $n$ qubits, which is equivalent to executing the composite gate $H^{\otimes n}$. Using the function \texttt{tensor\_n2} formalized in Definition~\ref{de1.5}, this gate can be constructed exactly.
\begin{lstlisting}[language=C,mathescape] 
 let n_hadamard = new_definition
  n_hadamard (n:$\mathbb{N}$) = tensor_n2 (cmatrix_list ($\lambda$k. hadamard) n)
\end{lstlisting}
Here, \texttt{cmatrix\_list} generates a list of n H gates, and \texttt{tensor\_n2} computes their tensor product. This operation transforms the initial state $|0\rangle^{\otimes n}$ into a uniform superposition.
\begin{equation}
\label{phi0}
    |\varphi_0\rangle=\frac{1}{\sqrt{N}}\sum_{x=0}^{N-1}|x\rangle
\end{equation}
\begin{lstlisting}[language=C,mathescape] 
 let zero_h = new_definition
  `zero_h = cmatrix_qstate_mul (hadamard_n:$\mathbb{C}^{2^N \times 2^N}$) (zero_qstate:(N)qstate)`
\end{lstlisting}
\textbf{(3) Construction of the Oracle Module $U_f$}

To mark the target item against all others, an Oracle operator is required. The function of this operator is to flip the phase of the target item, i.e., to apply a π-phase shift (by adding a negative sign).
\[
U_f|x\rangle = \begin{cases} 
|x\rangle,&\text{if } x\neq\tau \\
-|x\rangle,&\text{if } x=\tau
\end{cases}
\]
Here, the state $\tau$ is the solution state of the function; this phase transformation can be realized by a unitary matrix: $U_f=I-2|\tau\rangle\langle \tau|$.
\begin{lstlisting}[language=C,mathescape] 
 let U = new_definition
  `(U:$\mathbb{N}\rightarrow\mathbb{C}^{2^N \times 2^N}$) k = id_cmatrix - Cx(&2) %* project (tau k)`
\end{lstlisting}

The phase-flip operator tags the target state. This action is key to analyzing Grover's iterations. We formalize it as Theorem~\ref{de2.2}.
\begin{thm}
    \label{de2.2}
    Phase Flip
\end{thm}

The oracle operator $U_f$ performs a phase flip on the target state $|\tau\rangle$, satisfying for any coefficients $a,b\in \mathbb{C}$: $U_f(a|\tau\rangle+b|\tau^\perp\rangle) = b|\tau^\perp\rangle- a|\tau\rangle$.
\begin{lstlisting}[language=C,mathescape] 
 $\forall$k:$\mathbb{N}$ a b:$\mathbb{R}$. 1 $\leq$ k $\land$ k $\leq 2^N \implies$(U k:$\mathbb{C}^{2^N \times 2^N}$) ** (mk_qstate (Cx(a) %%% (tau k:(N)
qstate) + Cx(b) %%% (tau_perp k:(N)qstate))):(N)qstate = mk_qstate(Cx(b) %%% (tau_
perp k:(N)qstate) - Cx(a) %%% (tau k:(N)qstate))
\end{lstlisting}
Here, \texttt{\%\%\%} denotes the multiplication of a complex number with a quantum state. 
\\\\\textbf{(4) Construction of the Diffusion Operator $D$}

The diffusion operator $D$ performs an inversion about the mean on the amplitudes, which amplifies the probability amplitude of the target state. It is defined as $D = 2|\varphi_0\rangle\langle\varphi_0| - I$, with the corresponding matrix representation given below.
\[
D=\begin{pmatrix}
\frac{1}{2^{n-1}}-1 &\cdots &\frac{1}{2^{n-1}}\\
\vdots & \ddots &\vdots\\
\frac{1}{2^{n-1}} & \cdots &\frac{1}{2^{n-1}}-1
\end{pmatrix}    
\]

Since all evolution operations in quantum computing must be unitary, it is necessary to verify that the operator $D$ is a unitary matrix.
\\\\\textbf{(5) Construction of the Grover Iteration Operator}

The operator $G$ is composed of the Oracle and the diffusion operator: $G=DU_f=(2|\varphi_0\rangle\langle\varphi_0| - I)(I-2|\tau\rangle\langle \tau|)$. Since \( D \) and \( U_f \) have been verified as unitary matrices, their product is also a unitary matrix due to Theorem~\ref{de1.4}(Closure of Unitary Matrices Under Multiplication).

The operator $G$ is repeatedly applied to the initial superposition state $|\varphi_0\rangle$, with each application referred to as an iteration. To simplify the analysis, $|\varphi_0\rangle$ is decomposed into the target state $|\tau\rangle$ and the non-target state $|\tau^\perp\rangle$. By constructing a set of orthogonal bases, the entire quantum state can be confined to a two-dimensional subspace. Within this subspace, the initial state can be expressed as
   $ |\varphi_0\rangle=\sqrt{\frac{N-1}{N}}|\tau\rangle+\sqrt{\frac{1}{N}}|\tau^\perp\rangle$.
 Here, $|\tau^\perp\rangle=\sqrt{\frac{1}{N-1}}\sum_{x\neq x_0}^{N-1}|x\rangle$ represents the normalized superposition of all non-target states, where $N$ is the total number of states. To further simplify the expression, trigonometric functions are introduced to represent the coefficients:
\begin{equation}
\label{eq12}
|\varphi_0\rangle=cos(\theta)|\tau^\perp\rangle+sin(\theta)|\tau\rangle
\end{equation}
Where $\theta=arcsin(\sqrt{\frac{1}{N}})$.

\begin{figure}[h]
  \centering
  \includegraphics[width=0.8\textwidth]{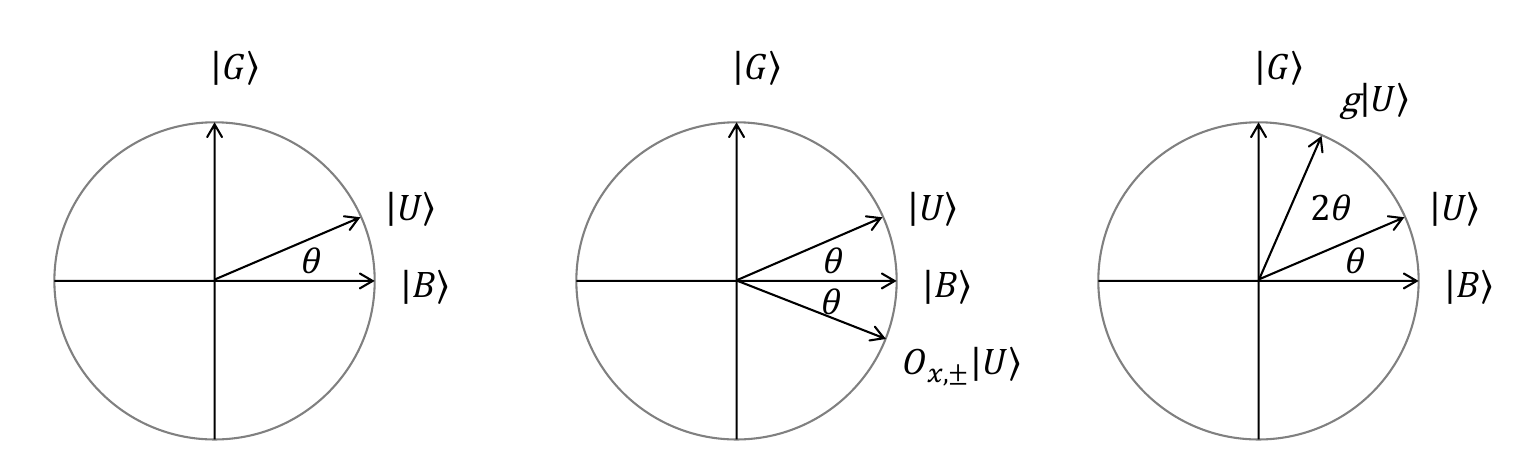}
  \caption{G operator iteration diagram}
  \label{figure2}
\end{figure}

The first iteration of Grover's algorithm starts from \( |U\rangle \) (the initial superposition state). Through a reflection about \( |B\rangle \) (the basis state) we obtain \( O_x, \pm|U\rangle \), then through a reflection about \( |U\rangle \) we obtain \( g|U\rangle \), gradually approaching the target state \( |G\rangle \). This transformation is represented geometrically in Figure \ref{figure2}.

After $t$ iterations, the state becomes:
\begin{equation}
\label{eq13}
|\varphi_t\rangle=G^t|\varphi_0\rangle=\cos\bigl((2t+1)\theta\bigr)|\tau^{\perp}\rangle
  +\sin\bigl((2t+1)\theta\bigr)|\tau\rangle
\end{equation}
Equation \ref{eq13} can be formally verified as the following theorem:
\begin{thm}
    \label{de2.3}
    Evolution under Grover Iterations
\end{thm}
\begin{lstlisting}[language=C,mathescape] 
$\forall$k t:num. 1 $\leq$ k $\land$ k $\leq 2^N \implies$ ((G k:$\mathbb{C}^{2^N \times 2^N}$) cmatrix_pow t) ** zero_h:(N)qstate
= mk_qstate(Cx(sin((&2 * &t + &1) * asn(&1 /sqrt(&$2^N$)))) %%% ((tau k):(N)qstate) 
+ Cx(cos((&2 * &t + &1) * asn(&1 / sqrt(&$2^N$))))%%% ((tau_perp k):(N)qstate))
\end{lstlisting}

Equation \ref{eq13} gives the explicit expression for the quantum state $|\varphi_t\rangle$ after $t$ iterations. To balance intuition with rigor, we use two complementary inductions to prove Theorem~\ref{de2.3}: algebraic and geometric. The algebraic approach works in the full $2^N$-dimensional Hilbert space, ensuring generality. The geometric approach uses the confinement of the dynamics to a two-dimensional subspace, which simplifies the proof. Together, they cross-validate the result and provide a multidimensional understanding of the Grover iteration mechanism.
\begin{itemize}
    \item \textbf{Algebraic induction}
\end{itemize}

We prove by mathematical induction that for all $t$, the identity $G^t|\varphi_0\rangle=\cos\bigl((2t+1)\theta\bigr)|\tau^{\perp}\rangle+\sin\bigl((2t+1)\theta\bigr)|\tau\rangle$ holds.

For the base case $t=0$, the identity becomes
\[
G^0|\varphi_0\rangle=cos(\theta)|\tau^\perp\rangle+sin(\theta)|\tau\rangle
\]
Since $G^0=I$, and the identity operator leaves any state vector unchanged, we have $
|\varphi_0\rangle=cos(\theta)|\tau^\perp\rangle+sin(\theta)|\tau\rangle$. According to Equation \ref{eq12}, this equality holds.

For the inductive step, assume the equation holds for some $t$, and prove that it also holds for $t+1$.
\\A1: $G^t|\varphi_0\rangle=\cos\bigl((2t+1)\theta\bigr)|\tau^{\perp}\rangle
  +\sin\bigl((2t+1)\theta\bigr)|\tau\rangle$
\\G1: $\implies$ $G^{t+1}|\varphi_0\rangle=\cos\bigl((2(t+1)+1)\theta\bigr)|\tau^{\perp}\rangle
  +\sin\bigl((2(t+1)+1)\theta\bigr)|\tau\rangle$
  
The proof is structured into four key stages: action decomposition, phase flip simplification, diffusion operator expansion, and trigonometric identity transformation. Each stage will now be elaborated in detail.
\\Step 1: Action Decomposition

First, using the definition of matrix exponentiation, we express $G^{t+1}|\varphi_0\rangle$ as $(G\cdot G^t)|\varphi_0\rangle$. Substituting the inductive hypothesis A1 directly, the left-hand side of the G1 equation becomes:
\[
    \label{eq16}
G\left(
  \cos((2t+1)\theta)|\tau^{\perp}\rangle
  +\sin((2t+1)\theta)|\tau\rangle
\right)  
\]
Then, using $G=DU_f$, we decompose a single application of the operator $G$. This application consists of the sequential action of the oracle operator $U_f$ followed by the operator $D$. The left-hand side now becomes:
\begin{equation}
\label{eq17}
 D\left(U_f\left(\,
\cos((2t+1)\theta)|\tau^{\perp}\rangle
+\sin((2t+1)\theta)|\tau\rangle
\,\right)\right)   
\end{equation}
Step 2: Phase Flip Simplification

According to Theorem~\ref{de2.2}(Phase Flip), Equation \ref{eq17} can be simplified to:
\begin{equation}
     \label{eq18}
D\left(\,
\cos((2t+1)\theta)|\tau^{\perp}\rangle
-\sin((2t+1)\theta)|\tau\rangle
\,\right)
\end{equation}
Step 3: Diffusion Operator Expansion

Substituting the operator $D=2|\varphi_0\rangle\langle\varphi_0|-I$ into Equation \ref{eq18} yields:
\begin{equation}
\label{eq19}
    \left(2|\varphi_0\rangle\langle\varphi_0| - I\right)\left(\!\,
\cos\bigl((2t+1)\theta\bigr)|\tau^{\perp}\rangle
-\sin\bigl((2t+1)\theta\bigr)|\tau\rangle
\,\right)
\end{equation}
Using the orthogonal normalization conditions \( \langle\tau^\perp|\tau\rangle= \langle\tau|\tau^\perp\rangle = 0 \) and \( \langle\tau|\tau\rangle = \langle\tau^\perp|\tau^\perp\rangle = 1 \), we compute the inner products:
\begin{equation}
    \label{eq21}
    \langle\varphi_0|\tau\rangle =  \cos\theta \langle\tau^\perp|\tau\rangle + \sin\theta \langle\tau|\tau\rangle = \sin\theta
\end{equation}
\begin{equation}
    \label{eq22}
    \langle\varphi_0|\tau^\perp\rangle =\cos\theta \langle\tau^\perp|\tau^\perp\rangle + \sin\theta \langle\tau|\tau^\perp\rangle = \cos\theta
\end{equation}
Substituting Equation \ref{eq21} and \ref{eq22} into Equation \ref{eq19} and expanding using the distributive property, we obtain
\begin{align}
& 2\cos\bigl((2t+1)\theta\bigr)\cos\theta \left( \cos\theta |\tau^\perp\rangle + \sin\theta |\tau\rangle \right) - 2\sin\bigl((2t+1)\theta\bigr)\sin\theta \left( \cos\theta |\tau^\perp\rangle + \sin\theta |\tau\rangle \right) \notag\\
& - \cos\bigl((2t+1)\theta\bigr) |\tau^\perp\rangle + \sin\bigl((2t+1)\theta\bigr) |\tau\rangle
\end{align}
Step 4: Trigonometric Identity Transformation

We combine the coefficients of \( |\tau^\perp\rangle \) and \( |\tau\rangle \) separately, yielding the following result:
\begin{align}
    \label{eq24}
& \Bigl( 2\cos\bigl((2t+1)\theta\bigr) \cos^2\theta - 2\sin\bigl((2t+1)\theta\bigr) \sin\theta \cos\theta - \cos\bigl((2t+1)\theta\bigr) \Bigr) |\tau^\perp\rangle\notag\\
&+ \Bigl( 2\cos\bigl((2t+1)\theta\bigr) \cos\theta \sin\theta - 2\sin\bigl((2t+1)\theta\bigr) \sin^2\theta + \sin\bigl((2t+1)\theta\bigr) \Bigr) |\tau\rangle
\end{align}

Then, using double-angle and sum-to-product trigonometric identities, Equation \ref{eq24} can be simplified to:
\begin{equation}
    \label{eq25}
\cos\bigl((2(t+1)+1)\theta\bigr) |\tau^\perp\rangle + \sin\bigl((2(t+1)+1)\theta\bigr) |\tau\rangle
\end{equation}
At this point, the left-hand side and the right-hand side of G1 are identical, completing the proof.
\begin{itemize}
    \item \textbf{Geometric induction}
\end{itemize}

Algebraic induction has verified the amplitude evolution formula after iterations of Grover's algorithm. In fact, this result can also be derived from a two-dimensional rotation perspective. In the two-dimensional space spanned by the target state \(|\tau\rangle \) and the non-target state \(|\tau^\perp\rangle \), the matrix form of the \( G \) operator in the \( \{ |\tau^\perp\rangle, |\tau\rangle \} \) basis is
$
\begin{bmatrix}
\cos(2\theta) & -\sin(2\theta) \\
\sin(2\theta) & \cos(2\theta)
\end{bmatrix}$.

The initial superposition state \( |\varphi_0\rangle = \cos\theta |\tau^\perp\rangle + \sin\theta |\tau\rangle \) has the following vector representation in the \( \{ |\tau^\perp\rangle, |\tau\rangle \} \) basis:
\begin{equation}
    \label{eq27}
    |\varphi_0\rangle = 
\begin{pmatrix}
\cos\theta \\
\sin\theta
\end{pmatrix}
\end{equation}

The vector form of the state after $t$ iterations of the $G$ operator, $G^t|\varphi_0\rangle = \cos\bigl((2t+1)\theta\bigr)|\tau^\perp\rangle + \sin\bigl((2t+1)\theta\bigr)|\tau\rangle$, in the $\{|\tau^\perp\rangle, |\tau\rangle\}$ basis is:
\begin{equation}
    \label{eq28}
    G^t|\varphi_0\rangle = \begin{pmatrix} \cos\bigl((2t+1)\theta\bigr) \\ \sin\bigl((2t+1)\theta\bigr) \end{pmatrix}
\end{equation}
To verify Equation \ref{eq28}, we apply the induction tactic, which splits the goal into two subgoals:
\\For the base case $t = 0$, we verify that $
    G^0|\varphi_0\rangle = \begin{pmatrix} \cos\theta \\ \sin\theta \end{pmatrix}$.
    
Since $G^0 = I$, and the identity operator leaves any state vector unchanged, we have $
    |\varphi_0\rangle = \begin{pmatrix} \cos\theta \\ \sin\theta \end{pmatrix}$.
According to Equation \ref{eq27}, this equality holds.
\\For the inductive step, assume the equation holds for some $t$, and prove that it also holds for $t+1$.
\\A1: $G^t|\varphi_0\rangle = \begin{pmatrix} \cos\bigl((2t+1)\theta\bigr) \\ \sin\bigl((2t+1)\theta\bigr) \end{pmatrix}$
\\G1: $\implies G^{t+1}|\varphi_0\rangle = \begin{pmatrix} \cos\bigl((2(t+1)+1)\theta\bigr) \\ \sin\bigl((2(t+1)+1)\theta\bigr) \end{pmatrix}$

The proof consists of three steps: inductive step, matrix multiplication, and trigonometric identity transformation. These steps are detailed below.
\\Step 1: Inductive Step

First, we express $G^{t+1}|\varphi_0\rangle$ as $(G \cdot G^t)|\varphi_0\rangle$. Then, by substituting the inductive hypothesis A1 and the rotation matrix form of the $G$ operator, the left-hand side of goal G1 becomes:
\begin{equation}
    \label{eq31}
    \begin{bmatrix}
\cos(2\theta) & -\sin(2\theta) \\
\sin(2\theta) & \cos(2\theta)
\end{bmatrix}
\begin{pmatrix}
\cos\bigl((2t+1)\theta\bigr) \\
\sin\bigl((2t+1)\theta\bigr)
\end{pmatrix}
\end{equation}
Step 2: Matrix Multiplication

After performing the matrix-vector multiplication, Equation \ref{eq31} becomes:
\begin{equation}
    \label{eq32}
    \begin{pmatrix}
\cos(2\theta)\cos\bigl((2t+1)\theta\bigr) - \sin(2\theta)\sin\bigl((2t+1)\theta\bigr) \\
\sin(2\theta)\cos\bigl((2t+1)\theta\bigr) + \cos(2\theta)\sin\bigl((2t+1)\theta\bigr)
\end{pmatrix}
\end{equation}
Step 3: Trigonometric Identity Transformation

Using the trigonometric sum-angle formulas, Equation \ref{eq32} simplifies to:
\begin{equation}
    \label{eq33}
    \begin{pmatrix}
\cos\bigl((2t+3)\theta\bigr) \\
\sin\bigl((2t+3)\theta\bigr)
\end{pmatrix}
\end{equation}
At this point, the goal is proven.

The two proof strategies complement each other. Algebraic induction relies solely on fundamental algebraic operations such as matrix multiplication and inner products, without appealing to any specific geometric intuition, thereby verifying the universality of the result in its most general setting. In contrast, geometric induction reveals the inherent two-dimensional rotation structure of Grover iterations. It shows that the exponential speedup stems from a geometric simplification of the dynamics in the effective subspace. Together, these methods not only cross-validate Theorem~\ref{de2.3} but also unify the algorithm's algebraic form with its geometric nature.
\\\textbf{(6) Measurement}

After \( t \) iterations, the probability amplitude of the target state \( |\tau\rangle \) reaches its maximum value. We then perform a projective measurement on the quantum state \( |\varphi_t\rangle \) at this moment, using the projection operator $
    P_{\tau}=|\tau\rangle\langle\tau|$.
Then, the probability of observing \( |\tau\rangle \) after measuring is thus 
\begin{equation}
\label{pt}
    p_t = \sin^2\bigl((2t+1)\theta\bigr)
\end{equation}

\section{Formal Verification of Grover's algorithm}
\label{section 4}

As mentioned in the previous section, the algorithm ends with a projective measurement to obtain the target solution with high probability. This probability is not fixed but varies with the number of Grover iterations $t$. To ensure efficiency, we formally analyze how the success probability $P_t$ varies with $t$. Specifically, we will verify its periodicity, optimality, and monotonicity in turn.

The success probability $p_t = \sin^2((2t+1)\theta)$ (Eq. \ref{pt}) exhibits a distinct periodic oscillation. Performance analysis must stay within one period. This implies that performance analysis of the algorithm must be confined to one complete period. determines the optimal number of iterations and prevents performance loss from over-iteration. Accordingly, the formal verification of this property is presented first.
\begin{thm}
    \label{de3.1}
    Periodicity of Probability
\end{thm}

The success probability \( p_t \) of Grover's algorithm is a periodic function with period \( \pi \). That is,
\begin{equation}
    \label{eq36}
    \sin^2\bigl((2t+1)\theta + \pi\bigr) = \sin^2\bigl((2t+1)\theta\bigr)
\end{equation}
\begin{lstlisting}[language=C,mathescape]
$\forall$x. sin(x + pi) pow 2 = sin(x) pow 2
\end{lstlisting}

The periodicity of the success probability \(p_t\) implies that the analysis of the algorithm can be confined to a single period. Within this interval, strict monotonic increase up to the peak guarantees that each Grover step systematically amplifies the solution amplitude. Verifying this amplification ensures the search converges deterministically rather than wandering randomly.
\begin{thm}
    \label{de3.2}
    Probability  Monotonic Increase
\end{thm}

For all iteration numbers $t$ such that $0<t \land t+1\le\frac{\pi}{4\theta}-\frac{1}{2}$, we have $0<p_{t+1}-p_{t}$ within the phase interval $[0,\tfrac{\pi}{2}]$.
\begin{lstlisting}[language=C,mathescape] 
 $\forall$t:$\mathbb{N}$. 1 $\leq$ k $\land$ k $\leq 2^N \land$ &0 < &t $\land$ &(t+1) $\leq$ pi / (&4 * asn (&1/sqrt(&$2^N$))) - 
 &1 / &2 $\implies$&0 < measurement (tau k:(N) qstate)(state_after_giter k (t + 1):(N)
 qstate) - measurement (tau k:(N)qstate) (state_after_giter k t:(N)qstate)
\end{lstlisting}
In Theorem~\ref{de3.2}, \texttt{state\_after\_giter} denotes the quantum state of the $G$ operator after $t$ iterations. We split the proof into two steps.\\
Step 1: Squared-difference to product

Substituting the success-probability formula gives $0<\sin^{2}\bigl((2t+3)\theta\bigr)-\sin^{2}\bigl((2t+1)\theta\bigr)$. Using the identity $\sin^{2}a-\sin^{2}b=\sin(a+b)\sin(a-b)$, we obtain $0<\sin\bigl((4t+4)\theta\bigr)\sin(2\theta)$.\\
Step 2: Sign of the factors

Given $\theta=\arcsin(1/\sqrt{N})$ and $0<t\land t+1\le\frac{\pi}{4\theta}-\frac{1}{2}$, we have
$0<\theta<\frac{\pi}{2}\quad$ and $0<(4t+4)\theta<\pi$.
Hence $\sin(2\theta)>0$ and $\sin\bigl((4t+4)\theta\bigr)>0$. The product of two positive factors yields $0<\sin\!\bigl((4t+4)\theta\bigr)\sin(2\theta)$,
completing the proof.

Following the probability amplification effect, we now examine its symmetric counterpart: probability decay. The monotonic decrease ensures that once the optimal iteration count is exceeded, performance declines predictably. Thus, it provides a theoretical basis for stopping iterations promptly and avoiding computational waste.
\begin{thm}
    \label{de3.3}
    Probability Monotonic Decrease
\end{thm}

Within the phase interval $[\tfrac{\pi}{2},\pi]$, for any iteration number $t$ satisfying $\tfrac{\pi}{4\theta}-\tfrac{1}{2}\le t \land t\le\tfrac{\pi}{2\theta}-\tfrac{3}{2}$, we have $0<p_{t}-p_{t+1}$. This proof follows the same lines as that of Theorem 4.2 and is therefore omitted.

\begin{lstlisting}[language=C,mathescape] 
 $\forall$t:$\mathbb{N}$. 1 $\leq$ k $\land$ k $\leq 2^N \land$ pi / (&4 * asn (&1 / sqrt(&$2^N$)) - &1 / &2 $\leq$ &t $\land$ 
 &t $\leq$ pi / (&2 * asn (&1 / sqrt(&$2^N$))) - &3 / &2 $\implies $&0 < measurement (tau k:(N)
 qstate)(state_after_giter k t:(N)qstate) - measurement (tau k:(N) qstate)(state_
 after_giter k (t+1):(N)qstate)
\end{lstlisting}

The foregoing properties establish a unimodal profile for $p_t$: monotonic increase on $[0,\pi/2]$ and decrease on $[\pi/2,\pi]$. Performance peaks when $(2t+1)\theta$ is closest to $\pi/2$. A strict and verifiable stopping criterion is therefore required. This criterion maps the theoretical optimum to an executable iteration count. It guarantees peak performance and eliminates resource waste from under- or over-iteration.
\begin{thm}
    \label{de3.4}
    Optimal Number of Iterations
\end{thm}

Within the interval $[0, \pi]$, to maximize \( p_t = \sin^2\bigl((2t+1)\theta\bigr) \), the phase angle must satisfy $t=\frac{\pi}{4\arcsin(1/ \sqrt{N})}-\frac{1}{2}$.
Since $t$ represents the number of iterations and must be a non-negative integer, there are only two practical values: the floor value $t_{\text{floor}} = \left\lfloor \frac{\pi}{4 \arcsin(1/\sqrt{N})} - \frac{1}{2} \right\rfloor$ and the ceiling value $t_{\text{ceil}} = \left\lceil \frac{\pi}{4 \arcsin(1/\sqrt{N})} - \frac{1}{2} \right\rceil$, i.e., $\forall t',\ 0 \leq (2t' + 1)\theta \leq \pi \implies p_{t'} \leq p_t \lor p_{t'} \leq p_{t+1}$. The monotonic increase (Theorem~\ref{de3.2}) and decrease (Theorem~\ref{de3.3}) help establish this optimal iteration count.
\begin{lstlisting}[language=C,mathescape] 
 $\forall$k t$'$:$\mathbb{N}$. 1 $\leq$ k $\land$ k $\leq 2^N \land$ (&2 * &t$'$ + &1) * asn (&1 / sqrt(&$2^N$)) $\leq$ pi $\implies$ 
 measurement(tau k:(N)qstate)(state_after_giter k t$'$:(N)qstate) $\leq$ 
 measurement(tau k:(N)qstate)(state_after_giter k (num_of_real(pi / (&4 * asn (&1 
 / sqrt(&$2^N$))) - &1 / &2)):(N)qstate) $\lor$ measurement(tau k:(N)qstate)(state_after_
 giter k t$'$:(N)qstate) $\leq$ measurement(tau k:(N)qstate)(state_after_giter k (num_of_
 real (pi / (&4 * asn (&1 / sqrt(&$2^N$))) - &1 / &2) + 1):(N)qstate)
\end{lstlisting}

This section builds on the state evolution model from section~\ref{section 3}. We formally verify Grover's algorithm's probability evolution. This reveals its complete dynamical characteristics.
Our findings show algorithm efficiency depends critically on precise iteration control. Within the finite phase interval bounded by periodicity (Theorem~\ref{de3.1}), the probability of success exhibits a unimodal characteristic of first monotonically increasing (Theorem~\ref{de3.2}) and then monotonically decreasing (Theorem~\ref{de3.3}) with iterations. The discrete optimal solution determination method (Theorem~\ref{de3.4}) realizes the effective connection between the theoretical optimal value and the actual iteration strategy. This series of properties collectively constitutes a rigorous theoretical foundation for the parameter optimization and efficiency evaluation of Grover's algorithm, providing a fundamental guarantee for its reliability and robustness in practical applications.
\section{Application}
\label{section 5}

In classical computing, integer factorization is a highly challenging problem, with computational complexity growing exponentially with the number of digits. Grover's algorithm can be used to accelerate this search process by exploring potential factors within the range $[2, \sqrt{N}]$ to find a non-trivial factor of $N$.

Taking the factorization of $143 = 11 \times 13$ as an example, it suffices to find one of the factors (either 11 or 13), as the other can be obtained via classical division. Since $\sqrt{143} < 12$, we take the closest power of 2 and use 4 qubits. Therefore, the search space size is $N = 2^4 = 16$. Our analysis will now focus on the optimal iteration count and the probability monotonicity for this scenario.
\\\textbf{1) Optimal Iteration Count for Integer Factorization}  

For the factorization instance with \(N = 16\), within the interval \([0, \pi]\), no success probability exceeds that of the floor or ceiling of the theoretical optimum. That is:
 \[
 \begin{aligned}[t]\begin{split}
 &\forall t.\ (2t + 1) \cdot \arcsin\left( \frac{1}{\sqrt{16}} \right) \leq \pi \implies p(t) \leq p\left( \left\lfloor \frac{\pi}{4\arcsin\left( 1/{\sqrt{16}} \right)} - \frac{1}{2} \right\rfloor \right) \\
& \lor p(t) \leq p\left( \left\lfloor \frac{\pi}{4\arcsin\left(1/{\sqrt{16}} \right)} - \frac{1}{2} \right\rfloor + 1 \right)
\end{split}\end{aligned}\]
Here, \(\lfloor \cdot \rfloor\) denotes the floor function. 
\begin{lstlisting}[language=C,mathescape] 
$\forall t:\mathbb{N}$.(&2 * &t + &1) * asn(&1 / sqrt(&$2^4$)) $\leq$ pi$\implies$ measurement(tau 11:(4)qstate)
(state_after_giter 11 t:(4)qstate)$\leq$ measurement(tau 11:(4)qstate (state_after_
giter 11(num_of_real (pi / (&4 * asn (&1 / sqrt(&$2^4$))) - &1 / &2)):(4)qstate) $\lor$ 
measurement(tau 11:(4)qstate)(state_after_giter 11 t:(4)qstate) $\leq$ measurement(
tau 11:(N)qstate)(state_after_giter 11(num_of_real (pi / (&4 * asn (&1 / 
sqrt(&$2^4$))) - &1 / &2) + 1):(4)qstate)
\end{lstlisting}
This conclusion can be verified by Theorem~\ref{de3.4}(optimal number of iterations ).
\\\textbf{2) Monotonicity Analysis of the Probability Function in Integer Factorization}  

\begin{figure}[ht]
 \centering
  \includegraphics[width=0.8\textwidth]{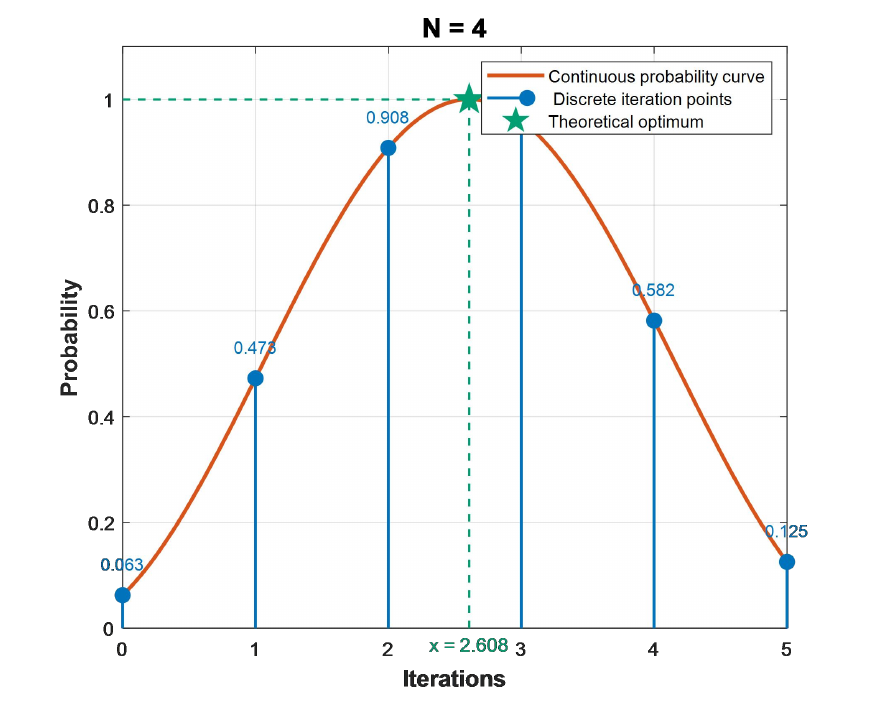}
  \caption{Non-monotonic success probability in Grover's algorithm}
  \label{figure3}
\end{figure}

Within one period (i.e., satisfying \((2t + 1) \cdot \arcsin\left( \frac{1}{\sqrt{16}} \right) \leq \pi\)), the valid range for the number of iterations is \(t \leq 5\). Within this range, the success probability \(p(t)\) exhibits a pattern of first increasing and then decreasing with the number of iterations, as illustrated in Figure \ref{figure3}.

We verify the increasing property before the optimal iteration, i.e., \( p(0) < p(1) \) and \( p(1) < p(2) \). This indicates that in the initial phase, the success probability increases monotonically with the number of iterations. The proofs of these two inequalities are similar. Taking \( p(0) < p(1) \) as an example, explain the proof process in detail.

\noindent Step 1: Inequality Transformation

Transform \( p(0) < p(1) \) into \( 0 < p(1) - p(0) \) to match the form of Theorem~\ref{de3.2}.

\noindent Step 2: Proving the Sufficient Condition

Given the theorem $P\Longrightarrow Q$, to prove the target \( Q \), we use the tactic \texttt{MATCH\_}
\texttt{MP\_TAC}, which replaces the goal with \( P \). Therefore, applying Theorem~\ref{de3.2}, the goal becomes proving:
\[
1 \leq \frac{\pi}{4 \arcsin\left( 1/{4} \right)} - \frac{1}{2}.
\]

\noindent Step 3: Trigonometric Inequality Scaling

Transform the target inequality into:
\[
\arcsin\left( \frac{1}{4} \right) < \frac{\pi}{6},
\]
and then, using the monotonicity of the inverse trigonometric function in the interval, convert it to proving:
\[
\frac{1}{4} < \sin\left( \frac{\pi}{6} \right).
\]

Next, we verify that after the optimal iteration, the success probability decreases monotonically with the number of iterations, i.e., \( p(3) > p(4) \) and \( p(4) > p(5) \). The proof process is similar to that of the increasing property, except that the deduction relies on Theorem~\ref{de3.3}.

The conclusion of this verification goes beyond simple simulation observations. It mathematically rules out the possibility of fluctuations or plateaus in the success probability curve before reaching the peak, confirming its strict monotonicity. This implies that any quantum circuit that follows the Grover recipe inevitably and deterministically improves its performance as the iteration count increases (up to the optimum). Our proof turns the conventional qualitative notions of algorithmic "validity" and "reliability" into quantitative, machine-verified guarantees.
\section{Conclusions}
\label{section 6}
This paper presents a formal model for constructing and verifying Grover's algorithm using the HOL Light theorem prover. Specifically, by formally defining core concepts such as quantum states, quantum gates, unitary transformations, and quantum measurements, we have established a novel quantum framework that provides a foundation for the formal verification of quantum computing. Within this framework, we have verified several key properties of Grover's algorithm and demonstrated its potential application in solving search-based problems, using integer factorization as an example.
This work not only extends the formal analysis of quantum algorithms but also provides a formal verification foundation for other application scenarios in quantum computing, such as quantum communication and quantum cryptography.
Future work will focus on extending the proposed framework, including exploring more quantum algorithms and further investigating their applications in more complex scenarios, coupled with in-depth verification incorporating quantum error-correcting codes and high-dimensional quantum systems.

\section*{Acknowledgment}
  \noindent This work was partly supported by the Natural Science Foundation of China (62372312, 62272323, 62272322, 62372311).

\bibliographystyle{alphaurl}
\bibliography{reference}

\end{document}